# Evolution of barchan dune interactions investigated by a downscaled water tunnel experiment: the temporal characteristics and a soliton-like behavior

**Nan He[1], Yang Zhang[1*], Yuanwei Lin[1], Bin Yang[2*], Xin Gao[3], Pan Jia[4]**

[1]Dept. of Fluid Machinery and Engineering, Xi'an Jiaotong University, Xi'an 710049, China.

[2]School of Chemical Engineering, Northwest University, Xi'an 710069, China.

[3]State Key Laboratory of Desert and Oasis Ecology, Xinjiang Institute of Ecology and Geography, Chinese Academy of Sciences, Urumqi 830011, China

[4]School of Science, Harbin Institute of Technology, Shenzhen 518055, China.

[*]Correspondence: zhangyang1899@mail.xjtu.edu.cn; binyang@nwu.edu.cn

**Key Points:**

- Characteristic time scales for the interactions of two barchan dunes are studied via a downscaled water dune experiment.

- A new pattern of dune interaction is found and regarded as an important critical state. Effects of the flow between the two dunes are described.

- An amazing soliton-like behavior of barchan dune is produced, and the factors affecting its survival are analyzed.

**Abstract**

This paper reports a downscaled water tunnel experiment to study the temporal characteristics of a double dune interaction system and the new pattern of dune interaction when the initial mass ratio of the two dunes is large. These topics are useful for a comprehensive understanding of the dune interaction system but were rarely covered before. The turnover time scale under dune interaction is defined, and its time averaged value is found to have a nonmonotonic relationship with the initial mass ratio. A nonmonotonic relationship is also found between the convexity of the downstream dune tip and the initial mass ratio. The stationary points of the two nonmonotonic curves above correspond to the same dune interaction pattern named "exchange-chasing," which is considered indispensable in the classification map of dune interactions. The upstream dune acts as an energy transmitter between fluid flow and the downstream dune. A soliton-like behavior occurs when the downstream dune enlarges, where a small dune is detached from the downstream dune tip and gets passed by the upstream dune approximately without mass exchange. The activity of such temporary soliton is found to be negatively related with the initial dune spacing and positively related with the initial mass ratio.

**Plain Language Summary**

The downscaled water tunnel experiment can achieve local similarity in sand dune morphology with the field results while avoiding the large spatiotemporal scales in field observations. A classic example is the solitary-wave-like behavior of dunes. A small dune merges in the upstream face of a large dune and pushes out a dune of similar size from the leeside of the large dune, which appears as if this small dune goes straight through the large one under monochromatic conditions. For this phenomenon, the existence in water tunnel experiments has been confirmed before the satellite image evidence was established. Aside from the solitary-wave-like behavior, researchers have discovered several different dune interaction patterns through water tunnel experiments, which are affected by various factors, such as flow rate and initial mass ratio of the two dunes. Assumptions show that when two dunes are close in size, they keep moving at the same celerity because the size of a dune is inversely proportional to its speed. In this paper, a new pattern emerges unexpectedly because of the flow structure between the two dunes. Specifically, the tip of the downstream dune is captured by the reversing flow produced by the upstream dune and disengages from the body of downstream one under the conditions of suitable mass ratio and spacing. Together with the upstream dune, this tip dune displays an amazing soliton-like behavior. Different from the abovementioned solitary-wave-like behavior, this soliton realizes the mutual crossing of dunes and the maintenance of self-mass at the same time within a certain space-time range. The reasons for its occurrence and the conditions affecting its survival are analyzed.

# 1 Introduction

Barchan dunes, which are crescent-shaped and highly mobile under predominantly unidirectional flow and limited sand supply, can be widely found in deserts, riverbeds, oil pipelines, and on diverse celestial bodies (Al-lababidi et al., 2012; Andreotti et al., 2002; Bagnold, 1941; Herrmann & Sauermann, 2000; Hersen et al., 2004; Parteli & Herrmann, 2007; Sauermann et al., 2000). A typical barchan dune shows two horns pointing downstream and a brink line dividing the dune body into windward and slip sides. Flow dragging sand grains goes along the windward surface and decelerates at the brink line due to the boundary layer separation, causing the deposition of sand grains in the slip face (Andreotti et al., 2002; Fischer et al., 2008; Hersen, 2004). In the recirculation zone, which is downstream of the slip face and between the two horns, a sand trap effect occurs, capturing grains from the slip face avalanche, with some of the remaining grains released from the horns. The repeating behavior of dragging, avalanche, and trapping leads to the stable morphology for a barchan dune when it migrates downstream (Génois et al., 2013). Therefore, the crescent shape is considered a strong attractor (Fischer et al., 2008; Groh et al., 2008; Alvarez & Franklin, 2018), which can be maintained at different spatiotemporal scales in accordance with diverse environments, varying from the centimeter and minute under water (Alvarez & Franklin, 2017; Endo et al., 2004; Hersen, 2005) to the kilometer and millennium on Mars (Claudin & Andreotti, 2006; Parteli & Herrmann, 2007).

The evolution of isolated barchan dune has been extensively studied (Alvarez & Franklin, 2017; Endo et al., 2004, 2005; Groh, 2008; Hastenrath, 1967; Hersen et al., 2002; Hersen, 2004, 2005; Zhang et al., 2014). Researchers have investigated the variations of the spatial characteristics of dune body, such as length, width, and the aspect ratio of dune bottom or longitudinal slice. Some studies have focused on the temporal characteristics, which are equally important for the characterization of dune evolution. As the barchan evolution can be recognized as a process-response system (Allen, 1974), it has two associated characteristic time scales, namely, the relaxation time and the turnover time (Hersen et al., 2004). When a disturbance is introduced in the system, the original equilibrium state of dune morphology is broken, and a new equilibrium cannot be reached until the relaxation time $t_r$ has been through (Allen, 1974; Parteli et al., 2009). The turnover time $t_o$ is the time needed for a dune to propagate over its own length, which can be expressed as $t_o = L/C$, where $L$ is the characteristic length, and $C$ is the migrating celerity. $L$ can be either the streamwise length or width (Andreotti et al., 2002; Groh et al., 2009), and $t_r$ is larger than $t_o$ (Hersen, 2004). In some studies, the barchan dune evolution has been characterized on the basis of $t_o$, such as the morphological variation of a dune field in an exposed water table (Luna et al., 2012) and the process of transverse dunes decaying into crescent shape (Parteli et al., 2011). A subaqueous experiment by Alvarez and Franklin (2017) proposed a relaxation period ($t_r = 2.5t_c$) for an isolated dune to evolve from a conical pile to its equilibrium state, that is, the crescent shape, in which $t_c$ is defined by incorporating several physical quantities.

$$t_c = \frac{L_{eq}}{C_{eq}} = \frac{L_{eq} \left( \rho_s / \rho \right) \left( \rho_s / \rho - 1 \right) g d}{\left( u_*^2 - u_{th}^2 \right)^{3/2}}, \tag{1}$$

where $L_{eq}$ and $C_{eq}$ are the characteristic length and celerity at equilibrium, respectively, $d$ is the sand grain diameter, $g$ is the acceleration of gravity, and $\rho$ and $\rho_s$ are the fluid and sand densities, respectively. $u_*$ and $u_{th}$ are the wall shear velocity and threshold velocity for sand grain entrainment, respectively. Here, $t_c$ denotes the value of $t_o$ at equilibrium.

Considering that dune fields (or corridors) are observed more frequently than the isolated case in nature, exploring the temporal characteristics of barchan dune evolution in the context of dune interactions is interesting. To the best of our knowledge, this topic has been rarely studied.

Therefore, dune interactions, which have been extensively investigated, are fundamental to advancing the research on such temporal issue. The dune interaction pattern known as "solitary-wave-like behavior" once sparked an interesting debate. In the beginning, Schwämmle and Herrmann (2003) found through simulation that barchan dune can behave as "soliton" or "solitary wave" under a high volume ratio between the two dunes, indicating that the upstream small dune "passes through" the downstream large one without changing their shapes. This idea has been challenged by a lack of geomorphological evidence showing small dunes emerging from the slip face of large ones. Livingstone (2005) considered that the small dune at the wake of a large dune may have migrated laterally from the sides into the wake zone owing to cross winds. Schwämmle and Herrmann (2005) then responded by citing the collision pattern "ejection" found in a water tunnel experiment by Endo et al. (2004) and pointed out that small dunes can calve from the slip side of large ones, and that barchans, behaving like a solitary wave, can retain their shape by a transfer of mass rather than by a transfer of momentum. However, Elbelrhiti et al. (2005) proposed that barchan dunes are inherently unstable and uneasy to exhibit solitary-wave-like behavior. The situation reversed again. Vermeesch (2011) investigated the satellite image sequences that include 10 dune collisions in the field for a time span of 45 years, showing clearly the evidence of solitary-wave-like behavior in the field, which was supported by the long-term field observation by Hugenholtz and Barchyn (2012). Therefore, the downscaled experiments can provide effectively analogical information from a morphological perspective while avoiding the large spatiotemporal scales in field observation despite the limited similarity to field conditions. The above story is depicted in Fig. 1, where the solitary-wave-like behavior produced in the present water tunnel experiment has high similarity in morphology with that in the field. In particular, it vividly demonstrates the mass exchange process using red and blue-dyed sand grains, where the sand grains that make up the small dune have been replaced after collision. As suggested by Schwämmle and Herrmann (2005) and Gu (2013), the small dune acts like a solitary wave by a transfer of mass, indicating that it cannot pass through the large one with its body intact like a real "soliton."

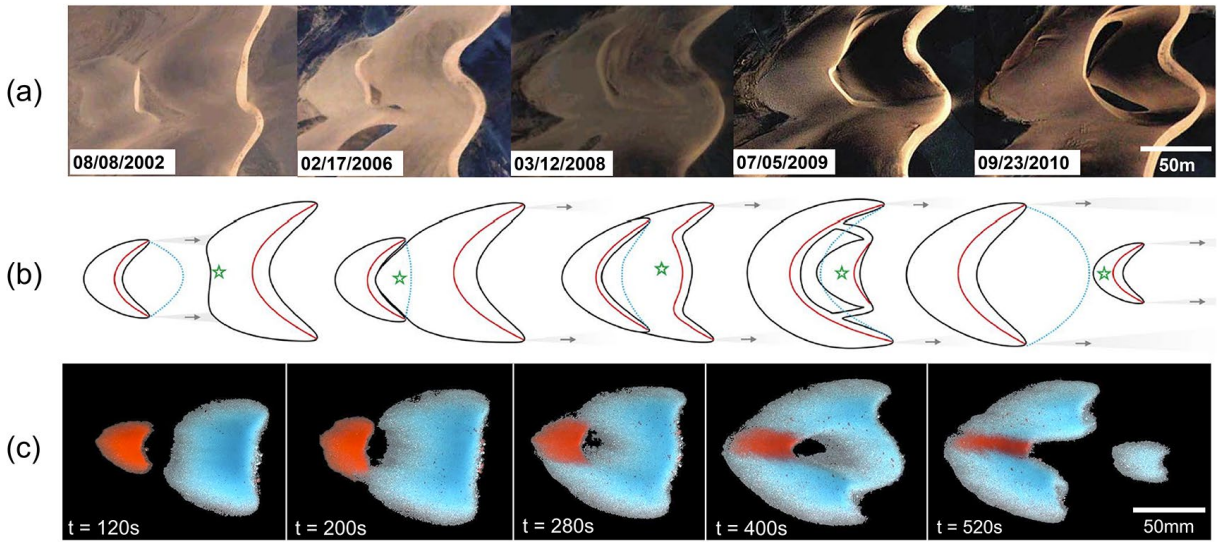

**Figure 1**. Solitary-wave-like behavior of dune collision (a) observed in nature, (b) illustrated for interpretation, and (c) realized in the present water tunnel experiment. Panels (a) and (b) come from Hugenholtz and Barchyn (2012).

To date, in most of the studies of dune interactions through downscaled experiments and scale-invariant simulations, the upstream dune is smaller than the downstream one, leading to reproducible interaction (or collision) patterns. Lima et al. (2002) presented a model involving the interdune sand flux and the collision pattern "coalescence" that refers to the pattern "merging" in Assis and Franklin (2020). A subaqueous experiment performed by Endo et al. (2004) showed that the inline collision between two dunes can be nominally described as a function of the mass ratio of the dunes, that is, different mass ratios will induce different collision processes, such as "absorption," "ejection," and "split," which correspond to "merging," "exchange," and "fragmentation-chasing" in Assis & Franklin (2020), respectively. Hersen and Douady (2005) explored the collision process of two dunes with different offsets by a double-color experiment, showing a mass exchanging process that occurs during collision. The collision patterns were investigated through simulation (Durán et al., 2009; Katsuki et al., 2005, 2011), in which the results are similar with the patterns "merging" and "exchange." On the basis of the collision patterns obtained in water tunnel experiments (Endo et al., 2004; Hersen & Douady, 2005), Génois et al. (2013) built an agent-based model and revealed the statistical properties of dune group that is highly selective in size distribution. Recently, a fundamental understanding of dune interaction patterns has been obtained through a downsized subaqueous experiment with explanations at grain scale (Assis & Franklin, 2020, 2021), where five collision patterns emerged with the initial mass ratio smaller than one. When the initial mass ratio exceeds 1, the expectant scenario might be that the small downstream dune migrates faster than the large upstream one because the celerity of an isolated barchan dune is inversely proportional to its size (Andreotti et al., 2002), and they move downstream together without contacting again, similar to the collision pattern named "chasing" (Assis & Franklin, 2020). However, no collision does not necessarily indicate "no physical interaction" or the fixed pattern of "chasing." Lima et al. (2002) and Génois et al. (2013) suggested that dunes with similar size can interact with each other without direct collision. Bristow et al. (2018, 2019) captured the flow field near the barchan surface by using particle image velocimetry measurements and presented that the energy generated by the turbulent wake downstream of the dune is important for dune interaction. Bacik et al. (2020, 2021) explored the quasi-two-dimensional aqueous dunes chasing each other in a narrow circular channel through experiments. They suggested that the turbulent flow structure at the wake of the downstream dune induces a repulsion effect that increases the migration rate of the downstream one, which is balanced by the mechanism that smaller dune moves faster, leading to the result that the two dunes migrate at the same celerity. Therefore, if the initial mass ratio is larger than one, then the interaction between the two dunes may show some new patterns on the basis of the knowledge in fluid mechanics that a larger obstacle will create a stronger and wider wake flow downstream.

To sum up, two issues must be explored: (1) characteristic time scales during dune interactions, and (2) patterns of dune interaction with the initial mass ratio larger than those in the published work. In this paper, a water tunnel experiment that incorporates double-dune interaction, three flow rates, several choices of the initial dune spacing, and a wide range of the initial mass ratio is conducted. The size of the downstream dune remains unchanged and is the same as that in an isolated dune evolution case. Thus, the evolution of the downstream dune under the influence of differently sized upstream dune can be studied by comparing with the isolated case. The rest of this paper is organized as follows. Section 2 presents the experiment. Section 3 discusses the two issues on the basis of the experimental results, where the discussion of the first issue naturally leads to that of the second. Section 4 provides the conclusion.

## 2 Experiment

The experiment is conducted in a closed water tunnel, which consists of a settling tank, four 1 m-long sections, a centrifugal pump, and a returning line, as shown in Fig. 2(a). The test section has a rectangular cross section with a height of 0.30 m and a width of 0.20 m. It starts at 2.6 m downstream of the inlet and 1.0 m upstream of the outlet. The tunnel is filled with water before the tests. The two conical piles are placed at two fixed locations and are aligned with each other along the flowing direction with a longitudinal separation that is 1.4 times as large as the bottom diameter of the downstream pile. The initial mass ratio of the two piles ($\xi_m$) and the mean flow rate ($V_f$) are the varying parameters in the present experiment. $\xi_m$ is defined as the mass of the upstream dune divided by that of the downstream one and it varies within [0.00, 1.70] by fixing the mass of the downstream dune at 10.0 g. The two dunes are made of round glass particles (0.15 mm $\leq d \leq$ 0.31 mm) with a density of $\rho_s$ = 3178 kg/m$^3$. Sand grains are divided and dyed with red and blue agents, which correspond to the two ends of the spectrum and are easily distinguished by image processing. The two piles pushed by the water flow at a constant rate are deformed and interact with each other, developing into crescent shapes. As shown in Fig. 2(b), the interaction process is recorded by a complementary metal–oxide–semiconductor camera with a spatial resolution of 1280×1024 pixel$^2$ at a frame rate of 50 Hz. A focal distance of 40 mm and a maximum aperture of F2.8 are applied. Two continuous light-emitting diode lamps are placed beside the test section to provide a homogenous and shadowless illumination.

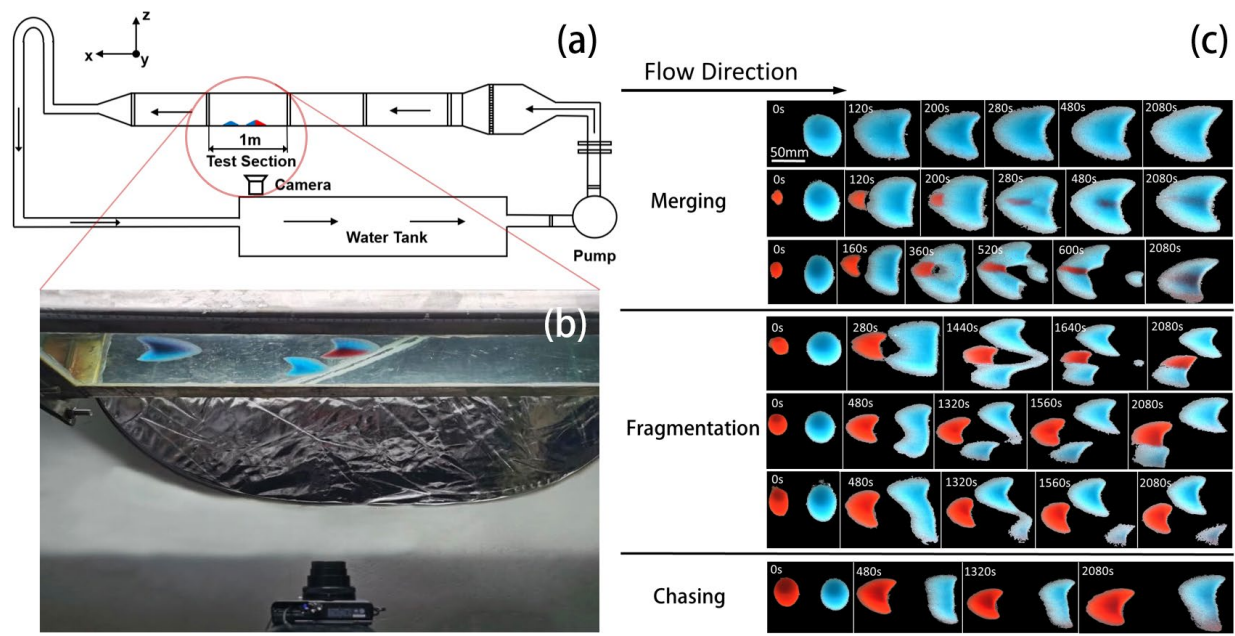

**Figure 2.** Water tunnel experiment of dune interaction. (a) Schematic of experimental setup; (b) Photograph of the test section that shows the high-speed digital camera, which is fixed on a traveling holder. The dune collision is recorded from the bottom view; (c) Snapshots from the bottom view of dune collision with the increase in the initial mass ratio ($V_f$ = 0.19 m/s). The patterns are named from top to bottom as the isolated case ($\xi_m$ = 0.00), merging ($\xi_m$ = 0.02),

exchange ($\xi_m = 0.05$), fragmentation-exchange ($\xi_m = 0.09$), exchange-chasing ($\xi_m = 0.23$), fragmentation-chasing ($\xi_m = 0.30$), and chasing ($\xi_m = 0.63$). The nomenclature here is consistent with that of Assis and Franklin (2020).

Fig. 2(c) shows the patterns that appear with the increase in $\xi_m$ (from 0.00 to 0.63). They can be divided into three categories designated as "merging," "fragmentation," and "chasing" by the behavior of the downstream dune. These patterns are reproduced successfully to what have been presented by Assis and Franklin (2020), except for a newly added pattern called "exchange-chasing" ($\xi_m \approx 0.23$), which is an indispensable pattern and important in describing the temporal features of dune interaction, as reported in the following section. The present work does not intend to repeat the published work because the cases corresponding to $\xi_m < 1$ are necessary to analyze the characteristic time scale under the dune interaction and to introduce the discussion with $\xi_m > 1$. The descriptions of these patterns can be found in the Supplementary.

## 3 Results and Discussions

For an isolated barchan dune, the turnover time $t_o$ is defined as its characteristic length divided by its celerity, where the characteristic length refers to the body length, width or the total length that includes the horn length. However, for the interaction cases in which the merging and fragmentation of dunes occur, another method is needed to properly define the characteristic length representing the interaction system. On a dune image corresponding to a certain moment, the connected domains can be classified into two types in accordance with their morphology, such as regular domain with crescent shape and irregular domain without crescent shape due to the connection of two or more crescent areas. As shown in Fig. 3, the moments of $t = 680$ s and 1080 s at $\xi_m = 0.15$ exemplify the latter type, where the crescent-shaped dunes connecting with each other are distinguishable. Re-emphasizing that the crescent shape, presented either separately or combined, is always the basic shape (and a strong attractor) in the dune interaction system is reasonable. Considering that a dune body can be treated as the coupling of a number of longitudinal slices (Andreotti et al., 2002; Groh et al., 2009; Kroy, et al., 2002), a single turnover cycle cannot be completed until the longest slice migrates over itself. Therefore, the instantaneous characteristic length of a connected domain can be defined as the length of the longest slice, which refers to the streamwise line segment crossing the most upstream point of this domain, as shown in Fig. 3. The instantaneous celerity of each domain can be easily obtained by an interframe tracking of the geometric centroid, which possesses a forward-difference scheme. At a moment, the instantaneous turnover time scale for a domain is its instantaneous characteristic length divided by its instantaneous celerity, and the mean value of the turnover time scales for all the domains represents the turnover time $t_o$ for the entire dune interaction system at this moment. Therefore, $t_o$ is a function of time, that is, $t_o = t_o(t)$, so it can be tracked continuously.

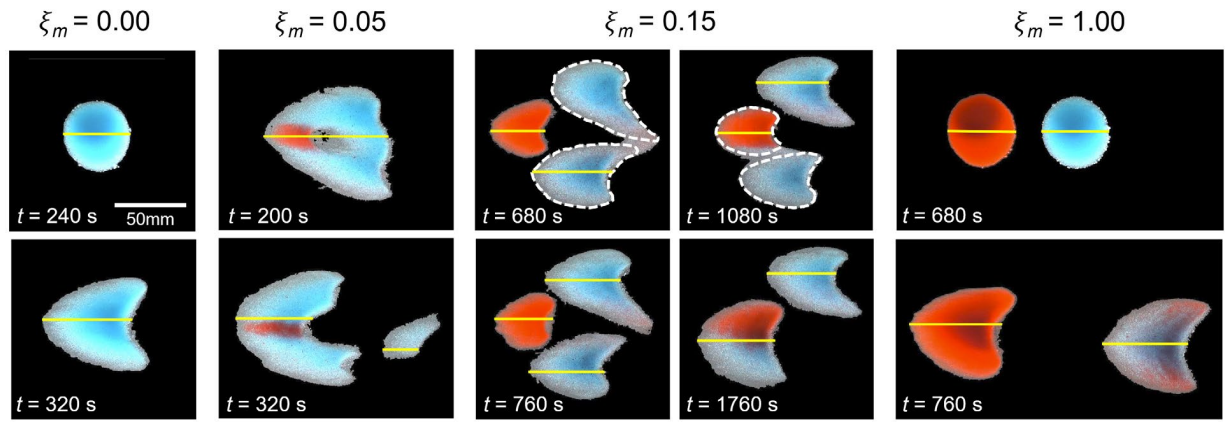

**Figure 3**. Examples of the characteristic length of dunes during interactions. The mean flow rate here is $V_f = 0.21$ m/s, and the flow direction is from left to right. The bright solid line segment along the flow direction denotes the characteristic length of each connected domain, which refers to the longitudinal slice crossing the most upstream point of the domain. The irregular domains

without typical crescent shape can be recognized as the connection by separated barchans that are enclosed with white dashed contours.

Fig. 4(a) shows an example of the variation of $t_o$ versus time, where the overall increase in its magnitude indicates a slowdown of dune evolution and a stabilized process of the interaction system (toward equilibrium) along with time. The fluctuation originates from the difference in characteristic length among the connected domains, such as the moment of $t$ = 320 s at $\xi_m$ = 0.05 in Fig. 3. Given that the small domain vanishes quickly, such fluctuation does not affect the abovementioned increase in $t_o$. The time-averaged value of $t_o$ for each run, named as $<t_o>$, is computed to explore the influence of the upstream dune (or $\xi_m$) on the temporal characteristics of the interaction pattern. The period for such time averaging needs to be determined ahead to compare the different cases of flow rates. Specifically, for $V_f$ = 0.19, 0.21, and 0.23 m/s, the period is settled as 2080, 1800, and 1200 s, respectively, which correspond to the same distance in the flow direction that the migrating dune system has covered. In addition, this distance roughly equals to the streamwise span of the present camera observation.

In an idealized situation where the upstream dune introduced in the system has no interaction with the downstream one, $<t_o>$ should increase with $\xi_m$ because larger barchan dune migrates slower (Andreotti et al., 2002). The real situation, which certainly involves dune interaction, exhibits otherwise. As shown in Fig. 4(b), a nonmonotonic relationship is found between $<t_o>$ and $\xi_m$, which is reproducible under three different flow rates. Specifically, at $\xi_m \approx$ 0.23 that corresponds to the newly introduced pattern "exchange-chasing," $<t_o>$ as a function of $\xi_m$ reaches the minimum (and also the stationary point on the curve). Given that the upstream dune is remarkably smaller than the downstream one, they inevitably collide with each other when $\xi_m <$ 0.23, and the increase in $\xi_m$ (or the size of the upstream dune) indicates stronger disruption that leads to a more scattered distribution of dunes. Thus, more dunes of small size and large celerity are generated, contributing more negatively to the mean value of $t_o$. Therefore, $<t_o>$ decreases with $\xi_m$. When $\xi_m > 0.23$, the upstream dune does not migrate as fast as in the previous cases, and the disruption it brings reduces. With the further increase in $\xi_m$, the interaction between the two dunes weakens and tends to the pattern "chasing", and then the consequence of increasing the mass of the upstream dune is similar to the ideal situation described above. Therefore, $<t_o>$ increases with $\xi_m$. To sum up, the pattern "exchange-chasing" corresponds to the most complex and disrupted situation for the dune interaction system. As a critical pattern that can be reproduced stably at different flow rates, it can be used to divide the interaction patterns into two categories in terms of temporal characteristics. Besides, $<t_o>$ decreases with the increase in $V_f$, indicating that the process of dune interaction becomes shortened (or more effectively done) by stronger flow energy.

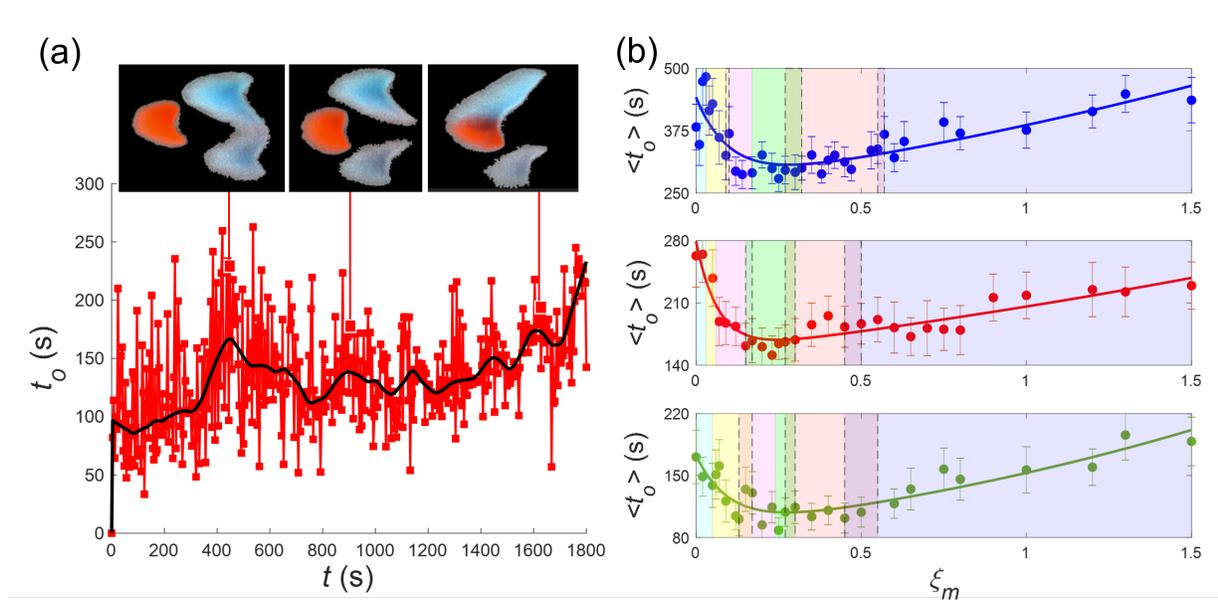

**Figure 4.** Turnover time scale for the dune interaction system. (a) Variation of the instantaneous turnover time $t_o$ with time ($\xi_m = 0.23$ and $V_f = 0.21$ m/s). The black line corresponds to the local averaging with an interval of 140 s, and three representative snapshots are illustrated by insets. (b) The time-averaged values of $t_o$ as a function of the initial mass ratio $\xi_m$ at three different flow rates

($V_f$ = 0.19, 0.21, and 0.23 m/s from top to bottom, which are represented by blue, red, and green, respectively) are fitted by a quadratic exponential based on the principle of least squares, and $R^2$ = 0.67, 0.84, and 0.79, respectively. The error bar for each value comes from the time-averaging of the data as in panel (a). The patterns "merging," "exchange," "fragmentation-exchange," "exchange-chasing," "fragmentation-chasing," and "chasing" are denoted by blue, yellow, pink, green, orange, purple patches from left to right, respectively, with the overlapped zones between the neighboring ones enclosed by dashed boxes.

As shown in Fig. 4(b), the value of $<t_o>$ for any dune interaction system with $\xi_m \neq 0$ is smaller than that of the isolated case ($\xi_m = 0$), except for a few cases ($V_f$ = 0.23 m/s and $\xi_m > 1.0$). This result indicates that the upstream dune, regarded as a disruptor, can accelerate the entire dune interaction system. The initial stage of dune interaction process is analyzed to explore in more detail how the flow energy is transferred through the upstream dune, in which the downstream dunes have not undergone severe deformation or fragmentation. Quantitatively, the initial stage corresponds to the period for the downstream dune to migrate over $d_0$, where $d_0$ is the equivalent diameter of its bottom at $t = 0$. The periods for $V_f$ = 0.19, 0.21, and 0.23 m/s are 240, 140, and 120 s, respectively. Fig. 5(a) takes the case of $V_f$ = 0.19 m/s as an example to illustrate the evolution of the downstream dune within the initial stage. With the increase in the mass of the upstream dune (or $\xi_m$), both the migrating distance and the degree of deformation toward crescent shape for the downstream dune increase, indicating that the upstream dune performs as an effective energy transmitter to provide the downstream dune with the kinetic energy and the energy committed to deformation work. Fig. 5(b) shows the variation of the mean celerity of the downstream dune during the initial stage, that is, $C_i$, with $\xi_m$ under different flow rates, where $C_i$ increases with $\xi_m$ in a convergent manner, indicating that the abovementioned energy transmission is balanced by some certain mechanism.

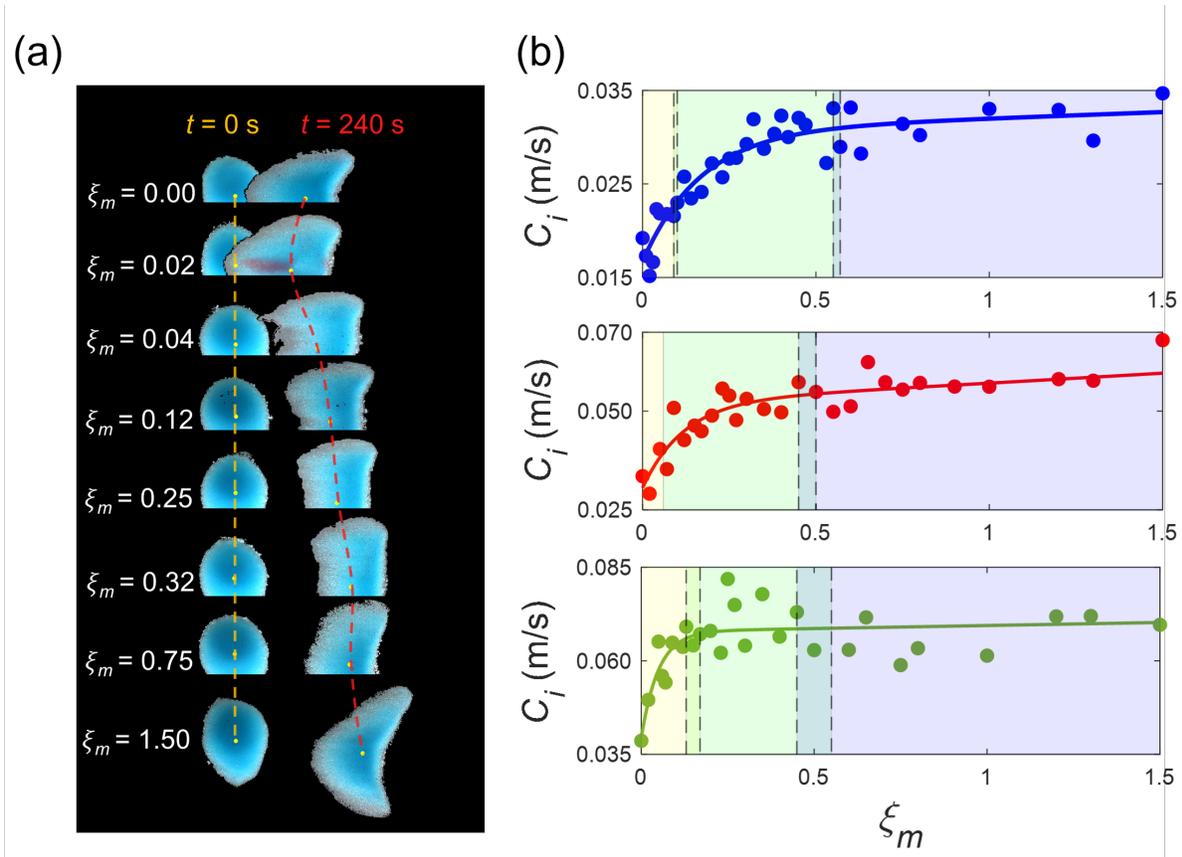

**Figure 5.** Evolution of the downstream dune during the initial stage. (a) Snapshots of the downstream dune at the beginning and end of the initial stage with the increase in the initial mass ratio ($V_f$ = 0.19 m/s). Dune centroids at the two moments are connected by the dashed yellow and red lines, respectively. (b) The variations of the mean celerity during the initial stage with the initial mass ratio under three different velocities ($V_f$ = 0.19, 0.21, and 0.23 m/s are denoted by blue, red, and green, respectively) are fitted by a quadratic exponential, and $R^2$ = 0.88, 0.82, and 0.60, respectively. The categories "merging," "fragmentation," and "chasing" are denoted by yellow,

green, and purple patches from left to right, respectively, with the overlapped zones between the neighboring ones enclosed by dashed boxes.

A qualitative explanation of this counteracting mechanism is given by virtue of the vorticity field around the two dunes, which is realized by a decoupled model simulation. The details of the simulation can be found in the Supplementary. The overview is shown in Fig. 6(a), where two rows of shedding vortices from the fringe of the upstream dune can be observed. The recirculation zone at the leeside of the upstream dune, which contains a pair of shedding vortices, plays as an intermediary that transmits the energy from flow to the downstream dune. In the initial stage, as the two dunes are sufficiently close to each other, the recirculation flow works as a spinning engine that deforms (the tip of) the downstream dune while propelling it in the flow direction. Such effect intensifies and the range of the recirculation zone enlarges when $\xi_m$ becomes larger, and the downstream dune undergoes a greater displacement, as shown in Fig. 5(a). In accordance with the comparison between Figs. 6(b) and 6(c), the reversing flow between the two shedding vortices expands with the increase in the relative size of the upstream dune and becomes larger than the width of the downstream dune. Thus, the trapping effect on the downstream dune gradually approaches (or even exceeds) the propelling effect, leading to the convergence of the celerity shown in Fig. 5(b). Therefore, the balance between the two effects can be considered a self-regulating mechanism, which guarantees that the increase in the size of the upstream dune cannot accelerate the downstream one unlimitedly. As shown in Fig. 5(b), the critical value of $\xi_m$ representing the convergence of $C_i$ appears earlier with the increase in flow rate, indicating that a stronger flow is more effective in strengthening the energy transmission between the two dunes. In that sense, the downstream dune obtains the same energy from a relatively smaller upstream dune. The patch representing the pattern "chasing" expands with the increase in flow rate, indicating that a stronger flow makes it difficult for the two dunes to directly contact with each other, especially when they are close in size.

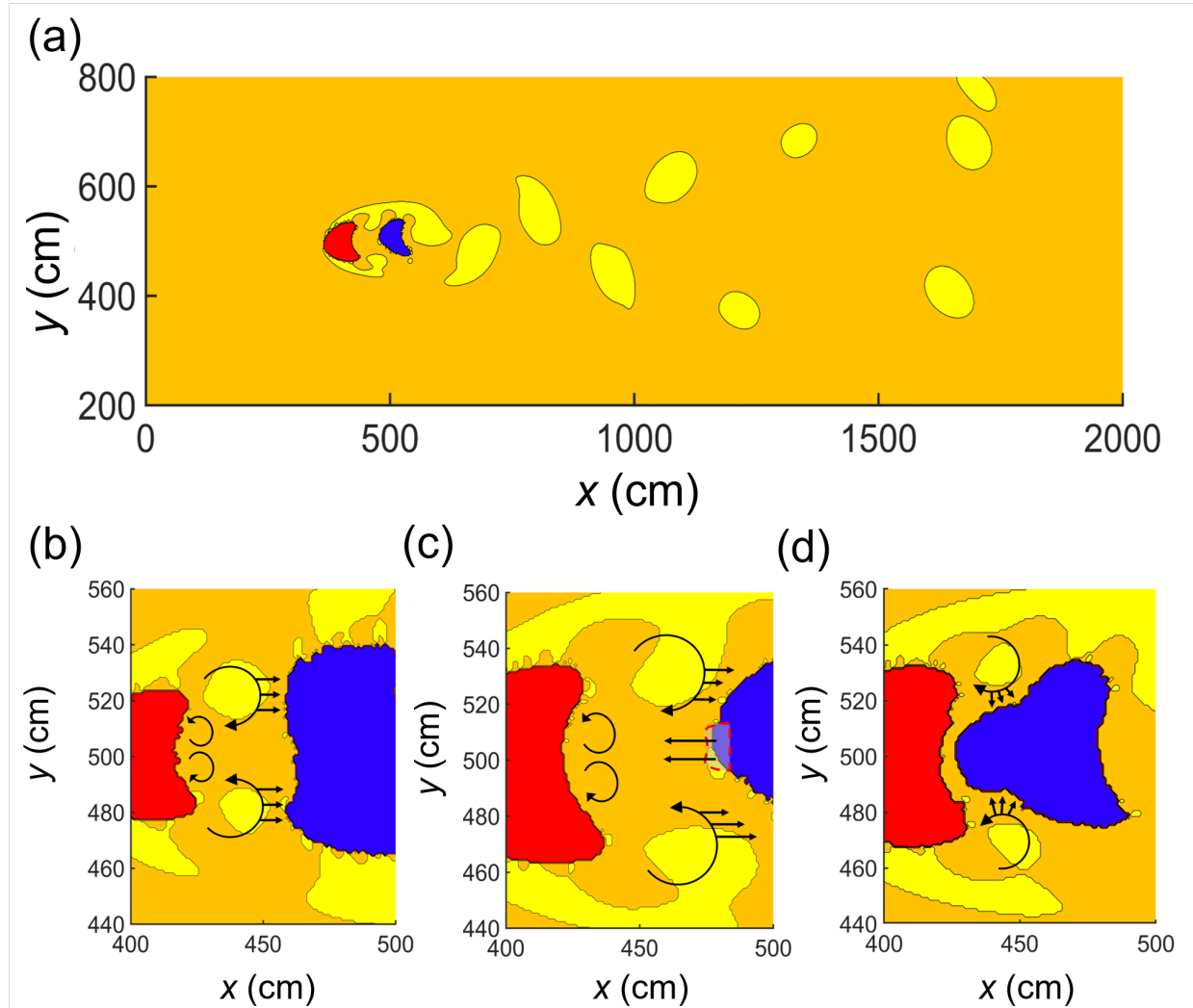

**Figure 6.** Simulated vorticity field between the two dunes based on a decoupled model and realized by Lattice Boltzmann method. (a) Overview, which is followed by the local views at (b) $\xi_m = 0.38$, $t = 240$ s, $V_f = 0.19$ m/s, (c) $\xi_m = 1.50$, $t = 240$ s, $V_f = 0.19$ m/s, and (d) $\xi_m = 1.70$, $t = 48$ s, $V_f = 0.24$ m/s. The vorticity field is binarized and marked in yellow and orange, and the upstream and downstream dunes are marked in red and blue, respectively. The circling and straight bold arrows in panels (b–d) denote the rotation of vortices and their effects. The small part of dune with a dashed contour in panel (c) indicates that the tip of the downstream dune is being pulled by the reversing flow.

With the increase in $\zeta_m$ (i.e., the size of the upstream dune), the recirculation flow between the two dunes expands and moves sideways, and the upstream rim of the downstream dune experiences a deforming process of first getting flattened and then getting sharpened, as illustrated by Figs. 6(b–d). A more detailed and quantitative description is shown in Fig. 7. The upstream surface of a barchan dune can be regarded as the superposition of a group of parabolas along the vertical direction, where the range enclosed by the parabola decreases with the increase in height (Sauermann et al., 2000). Therefore, the upstream rim of the crescent connected domain representing the downstream dune is fitted by a parabola, and the coefficient of the quadratic term (named *a*) is defined as the convexity of the dune tip, as shown in Fig. 7(a). The variation of the convexity at the final moment of the initial stage with $\zeta_m$ is shown in Fig. 7(b), where the abovementioned deforming process is quantified. The increase in flow rate increases the convexity, confirming to the perception that dune interaction intensifies with greater flow energy, as shown in Figs. 4(b) and 5(b). The set point in the curve of $a = a(\zeta_m)$ shows up stably for three different flow rates and corresponds again to the abovementioned pattern "exchange-chasing" ($\zeta_m \approx 0.23$). In other words, given a fixed initial dune spacing, this pattern can be regarded as the critical state that determines whether the upstream dune flattens or sharpens the downstream one.

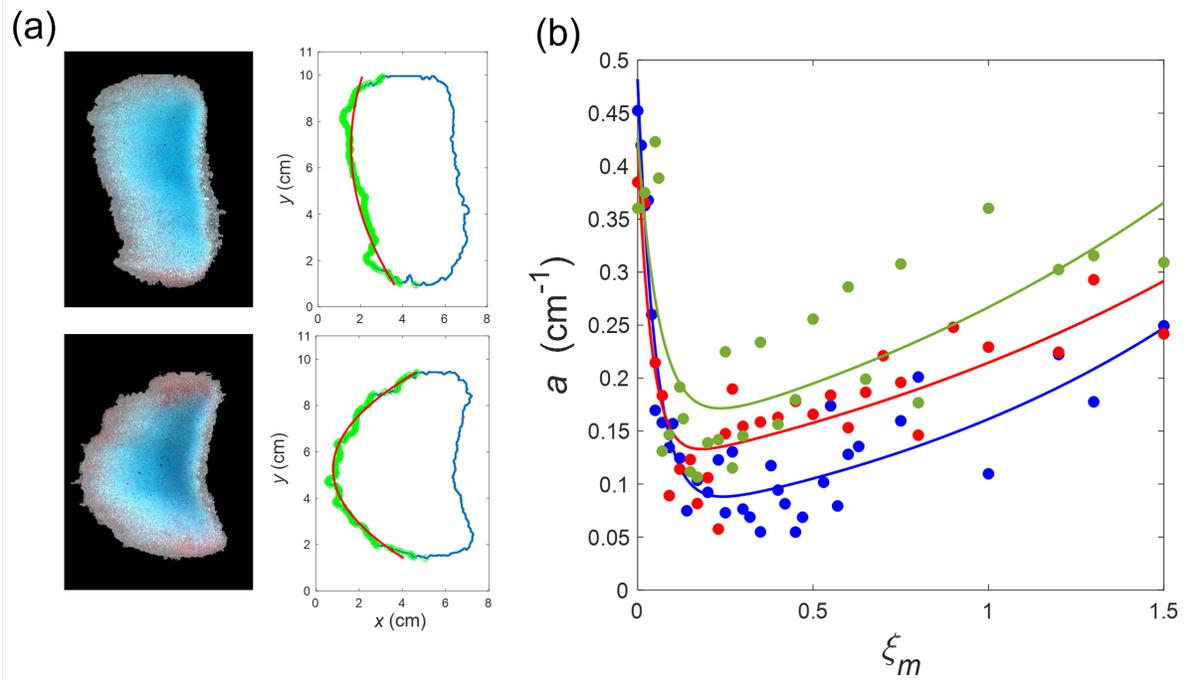

**Figure 7.** Convexity of the tip of the downstream dune at the final moment of the initial stage. (a) Computation of the convexity. Snapshots at $\xi_m = 0.23$ (top) and $\xi_m = 0.70$ (bottom) are shown as examples, where $V_f = 0.21$ m/s. The upstream and downstream rims are denoted by green and blue, respectively, and the quadratic fitting ($x = ay^2 + by + c$) is denoted by red, where $a$ indicates the convexity of the downstream dune tip. (b) The variations of $a$ with the initial mass ratio $\xi_m$ under three different flow rates ($V_f = 0.19$, 0.21, and 0.23 m/s are denoted by blue, red, and green, respectively) are fitted by a quadratic exponential, and $R^2 = 0.90$, 0.77, and 0.52, respectively.

Aside from the enhancement from the greater flow energy, the convexity of the downstream dune tip becomes increasingly larger and does not converge to a certain value when $\xi_m$ continues to increase, as shown in Fig. 7(b). This condition is caused by the trapping effect described above. However, focusing on the specific implementation of the trapping effect is necessary. As shown in Fig. 6(d), the reversing flow from two sides pulls the entire tip upstream like strangling a neck, leading to an increase in the convexity. Therefore, the downstream dune tip is likely to fall off the body rather than the downstream dune having an increasingly sharp crescent shape with the progress of the dune interaction through the initial stage and its further development.

The entire dune interaction process at larger value of $\xi_m$ and greater flow energy is studied to confirm this conjecture. As shown in Fig. 8, the reversing flow pulls the tip of the downstream dune and continuously enlarges its convexity during the initial stage ($t < 440$ s). At a certain moment, the tip is successfully disconnected from the downstream dune body. This disconnected part in the next stage suddenly exhibits a literally solitary behavior or one can tell that it becomes a "soliton." Specifically, the soliton stops in place after becoming separated and lets the upstream dune run over its body. In accordance with the principle of relative motion, the soliton passes through the upstream dune, and the sand composition of the two remains unchanged. We speculate that this interesting scene are due to several reasons: (1) the soliton formed by pulling is naturally a small and flat dune, which will not cause significant streamline deformation or flow separation, and the upstream dune stands between it and the mainstream flow, so that it may avoid the flushing by the reversing or the mainstream flow; (2) the upstream dune can easily go over the soliton surface, which is a gentle slope (the gentler the slope, the less likely the two dunes collide and cause sand mixing); (3) a sufficiently strong trapping effect produces a soliton with adequate mass, which can retain its basic shape and most of its mass after "rubbing" against the upstream red dune. At $t = 440$ s, as the original upstream dune passes through the soliton, a part of its sand grains is disengaged and merges with the soliton. This condition is because when the soliton comes into contact with the tip of the original upstream dune, a shape like a "funnel waist" is formed, which induces again the trapping effect (the red dune here plays the role of the blue downstream dune at $\xi_m < 1$). At $t = 560$ s, this soliton turns into the most upstream dune, which faces directly with the mainstream flow and deforms quickly into a crescent-shaped dune (or the soliton "dies"). During $t$ = 672–1088 s, this soliton triggers two consecutive collisions with the original upstream and downstream dunes. For the two collisions, $\xi_m \ll 1$, and the pattern is clearly "exchange" or "solitary-wave-like behavior." Obviously, this new "upstream dune" is extremely small to give birth to any new soliton in the system.

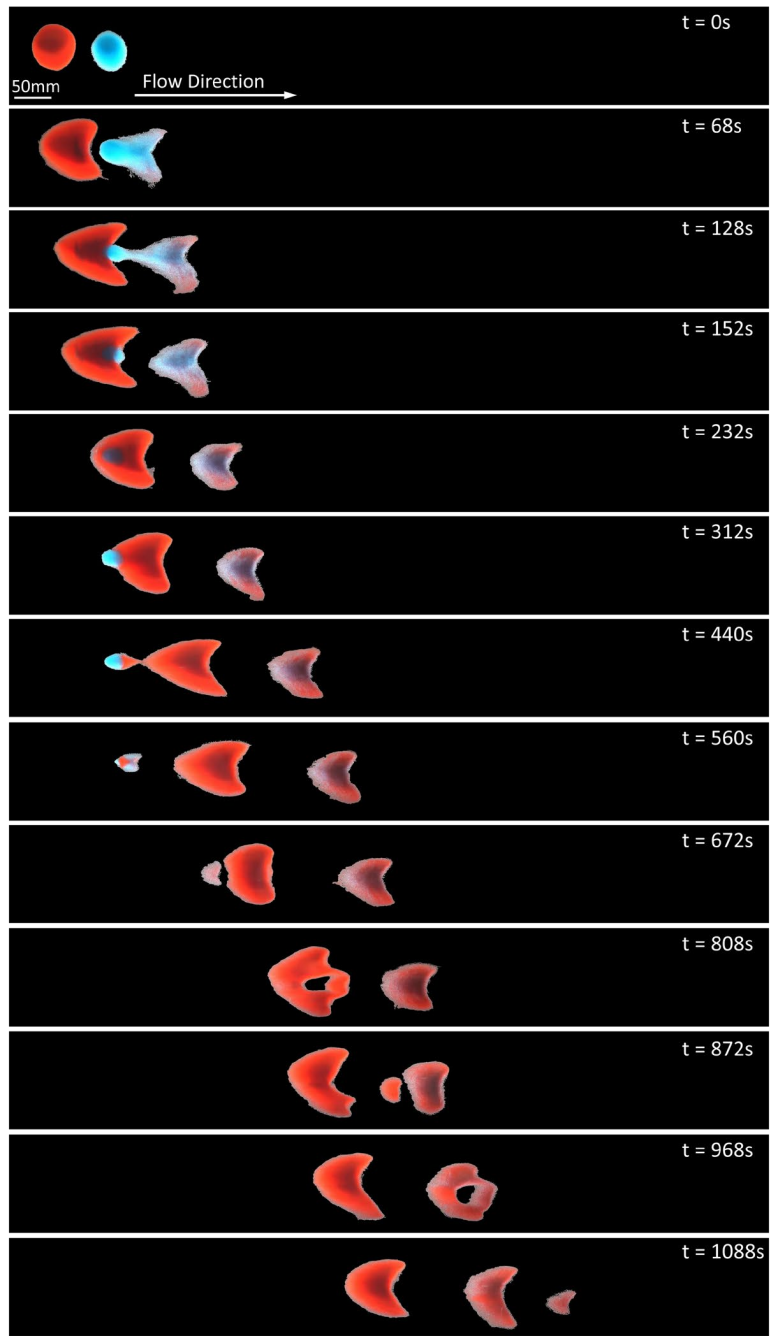

**Figure 8.** Collision case ($\xi_m$ = 1.7, $V_f$ = 0.24 m/s) that verifies the trapping effect of the upstream dune. Two consecutive "exchange" patterns demonstrate that the double-dune interaction can be regarded as the elementary process within a dune field evolution.

From Fig. 8, the size of the reversing flow, which is limited by the relative size of the upstream and downstream dunes, is key to the vigorousness of a soliton. Here, the vigorousness refers to the capability for a soliton to maintain the integrity of its own body. The initial streamwise spacing between the two dunes is also an important factor that cannot be ignored (Assis & Franklin, 2020). We investigate the vigorousness (the activity or the "lifespan") of soliton with the increase in the initial dune spacing, which is named $D_0$. As shown in Fig. 9(a), $D_0$ = 6.07 cm leads to a soliton behavior, as shown in Figs. 8 and 9(b) ($D_0$ = 6.63 cm), which is followed by two consecutive "exchange" patterns, indicating that the soliton behavior is reproducible. With a smaller $D_0$, the soliton in Fig. 9(a) appears to be more vigorous than that in Fig. 9(b) because it captures more red sand grains by friction when it "passes through" the tip of the upstream dune. As shown in Fig. 9(c), the reversing flow cannot pull out blue sand grains as many as those in the previous two conditions when $D_0$ increases to 7.39 cm. Accordingly, the soliton, which is insufficiently vigorous, fails to pass through the upstream red dune and then gets swallowed ($t$ = 316 s). This "internal" soliton, although swallowed, still works in an implosive manner that triggers the first "exchange" collision. A dune with small convexity is pushed out ($t$ = 388–472 s), which collides with the most downstream dune and realizes the second "exchange" collision ($t$ = 612–1220 s). As shown in Fig. 9(d), a similar process occurs when $D_0$ further increases to 8.17 cm, in which the soliton is less vigorous than that in Fig. 9(c) because it looks more difficult to pass through the upstream red dune, and the dune it pushes out by imploding is smaller. As shown in Fig. 9(e), the distance between the two dunes is extremely far to realize an effective trapping effect as $D_0$ becomes 9.00 cm. Accordingly, the soliton only appears in its rudimentary form ($t = 112$ s), which does not break away from the downstream blue dune. Thus, the two dunes migrate downstream at a similar celerity, forming a "chasing" pattern. For the vigorousness of soliton, the relative size of the upstream and downstream dunes determines its upper limit, and the initial dune spacing determines its lower limit (or the discount).

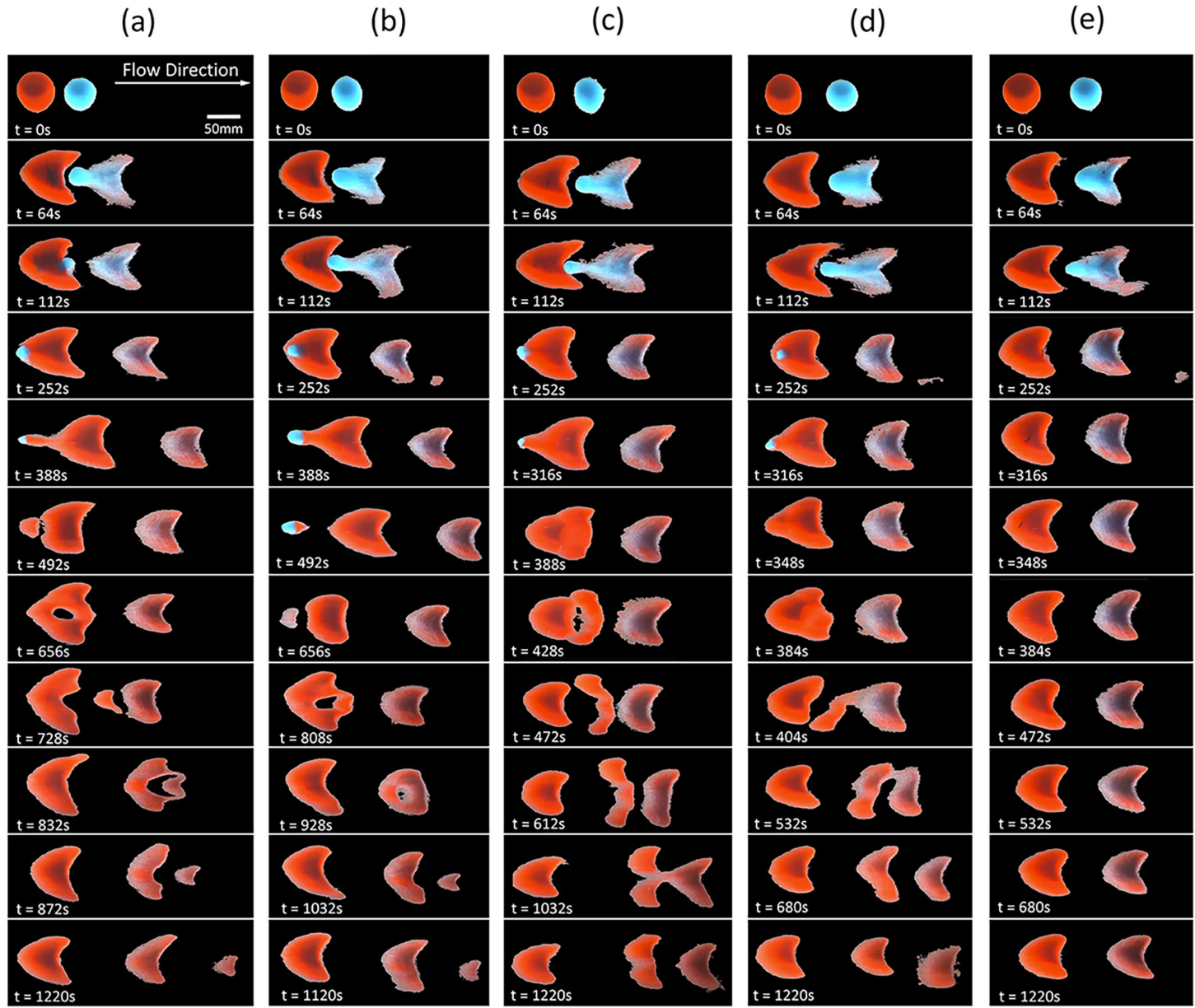

**Figure 9.** Snapshots of dune collisions ($\xi_m = 1.7$) with the increase in the initial streamwise distance between the two dunes $D_0$. (a) $D_0 = 6.07$ cm, (b) $D_0 = 6.63$ cm, (c) $D_0 = 7.39$ cm, (d) $D_0$ = 8.17 cm, and (e) $D_0 = 9.00$ cm.

## 4 Conclusion

This paper reports a water tunnel experiment to study the downscaled dune interactions, which covered different flow rates, initial dune spacing, and the widest range of initial dune mass ratio. The main points are summarized as follows.

First, the nonmonotonic relationship between the averaged turnover time scale and the initial mass ratio and its negative correlation with the flow rate are obtained. The critical state in this nonmonotonic relationship corresponds to a new interaction pattern called "exchange-chasing," which can be reproduced stably under different flow rates. When the initial dune spacing and flow rate are constant, this pattern also corresponds to another critical state, that is, whether the downstream dune goes toward crescent shape under the action of the upstream dune acting like an "energy transmitter." Second, the reversing flow between the dunes can cause two effects of "trapping" and "propelling," and their relative strength reflects the role of the "energy transmitter" of the upstream dune. Third, the so-called soliton behavior is obtained in studying the continuous sharpening of the downstream dune, that is, the two dunes maintain their respective shape after crossing each other without transfer of mass, which is in contrast to the observed solitary-wave-like behavior. Finally, the effects of the main factors on soliton behavior are investigated. The results show that such spacing determines the lower limit of soliton vigorousness, and the initial dune mass ratio determines the upper limit.

In future work, for the characteristic time scale, whether the dune interaction patterns will become "stable" at given conditions remains to be explored. On the one hand, one can expand the space for dune evolution by modifying the experiment, such as lengthening the straight water tunnel or adopting an annular water tunnel (where the periodic boundary conditions should be set carefully). On the other hand, one can define the "stability" of the dune interaction system from other perspectives by keeping the current experiment unchanged. The above methods are proposed to define the relaxation time scale of the dune interaction system, so that one can study the influence of main factors on it and establish the coupling relationship between different characteristic time scales, thereby enriching the understanding of the temporal characteristics of dune interaction process. For the soliton phenomenon, the parametric characterization of soliton activity needs to be realized, and experiments involving more different conditions are required. In this way, the comprehensive effects of factors, such as initial dune spacing, initial mass ratio, and flow rate on soliton behavior can be explored, thereby establishing the scaling law of soliton behavior. In the Introduction, we referred to the research process of solitary-wave-like behavior, in which evidence from downscaled experiments appeared earlier than that from field observations. This process demonstrates the significance of downscaled experiments in studying the morphological evolution of dunes (although a complete similarity is unlikely to happen) and makes us look forward to the follow-up of the current experiment: will anyone observe the soliton behavior of crescent dunes in nature? The condition conducive to soliton-like behavior should be difficult to achieve in nature, which requires that the upstream dune is suitably larger than the downstream one, and the spacing is right. However, it may be achievable if small dunes are blown by crosswinds into the recirculation zone downstream of large ones, as suggested by Livingstone (2005).

**Acknowledgments**

Yang Zhang is supported by NSFC (11402190), and Bin Yang is supported by KRDPSP (2021GY-150) and WSRCSP (XM06190109).

**Supplementary**

**Experiment**

The experiment was conducted in a water tunnel with transparent material, in which the temperature of water is around 18.5 ℃. The mean flow rate $V_f$ was set as 0.19 m/s, 0.21 m/s and 0.23 m/s, respectively. As shown in Fig. S1 and S2, $V_f$ and the corresponding friction velocity $u_*$ were computed based on the PIV measurements. Fig. S3 shows the average migration celerity of an isolated dune at different flow rates. Such isolated case can be used as the base (benchmark) of the dune interaction system. By comparing the interaction cases (with an upstream dune of non-zero mass) with the base case, one can find the influence of the upstream dune on the downstream one from the dynamical or morphological perspective. As shown in Fig. S4, the interaction between the dunes presents a set of patterns. Based on the behavior of the downstream dunes, these patterns are designated as "merging," "fragmentation," and "chasing," which can be further divided into seven sub-patterns. (a) "Merging" happens when the mass of the upstream dune is small (or even zero). As the upstream dune does not exist, the pattern is named "isolated case." As a small upstream dune exists, the two dunes merge into a mixed one after collision, indicating that the downstream dune remains complete. This pattern is named "merging." For a slightly larger upstream dune, when it moves downstream and reaches the downstream dune, a small blue dune is ejected from the mixed dune, which migrates downstream and eventually vanishes due to its size smaller than the minimum threshold. This pattern is classified as merging and specified as "exchange". (b) "Fragmentation" happens when the flow structure downstream of the red dune separates the blue dune. "Fragmentation-exchange" refers to the case in which the upstream dune is smaller and faster than the splitting downstream dune, causing an off-center mass exchange. As the mass of the upstream dune further increases, its leeside flow structure expands to make the downstream dune split before getting touched, and then the upstream dune catches up with one part of the downstream dune and merges with it. This pattern is specifically designated as "chasing-exchange." When the upstream dune becomes sufficiently large, "fragmentation chasing" happens, in which the upstream dune cannot reach the split downstream dunes, as bigger barchans migrate slower. (c) "Chasing" happens when the upstream dune becomes too big and completely unable to split or catch up with the downstream dune. Information for all the tests is included in Tab. S1.

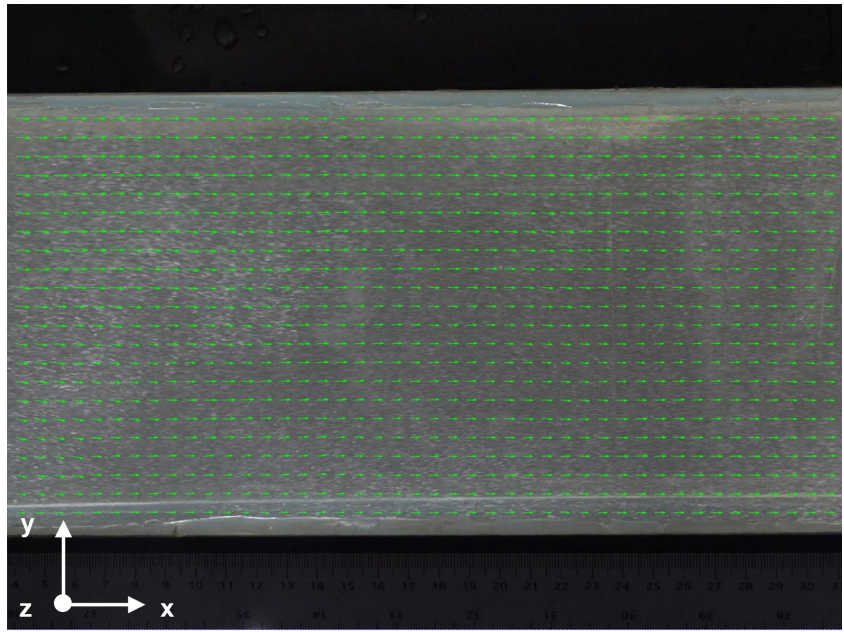

**Figure S1.** Flow field measurements of the test section from bottom view using PIV correlation algorithm before the dunes are put in. The upper and lower boundaries of image represent the sidewalls of water tunnel.

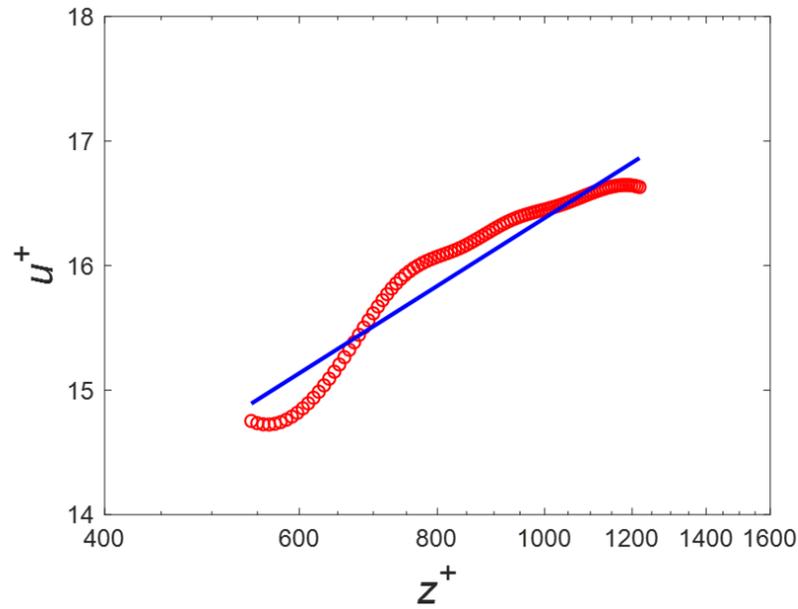

**Figure S2**. Flow velocity profile obtained from side view. The fitting result (in blue line) to the spatial-averaged scattered data (in red circle) is in accordance with the logarithmic law $u^+ = 1/\kappa \cdot \ln(z^+)+C$, where $u^+ = u/u_*$ is the normalized velocity, $\kappa = 0.41$ is the von Kármán constant, $z^+ = zu_*/v$ is the normalized height, $C$ is a constant, $u = u(z)$ is the mean velocity of the fluid, $u_* = 0.01267$ m/s is the shear velocity, and $v$ is the kinematic viscosity of fluid.

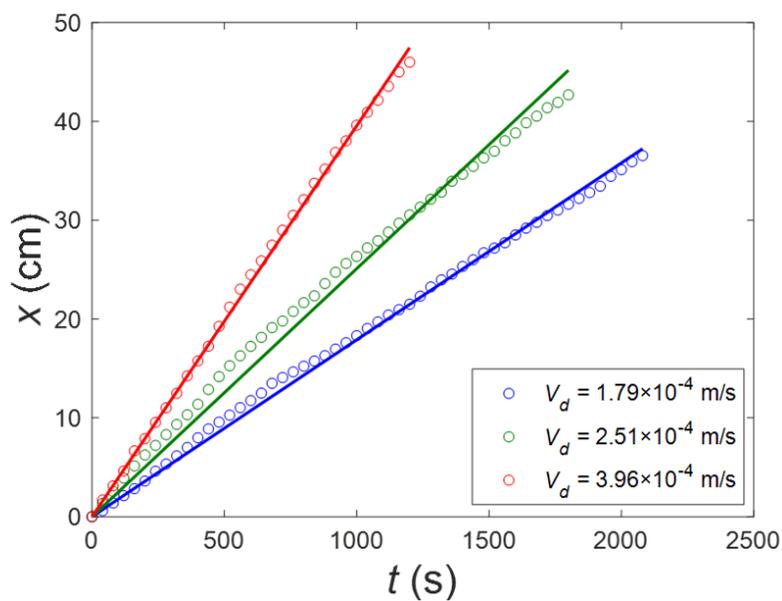

**Figure S3.** Displacement of an isolated dune under different flow rates. $V_f$ = 0.19 m/s, 0.21 m/s and 0.23 m/s are represented by blue, green and red, respectively. The solid lines are the corresponding liner fitting results, of which the slope is the average celerity of the isolated dune, $V_d$.

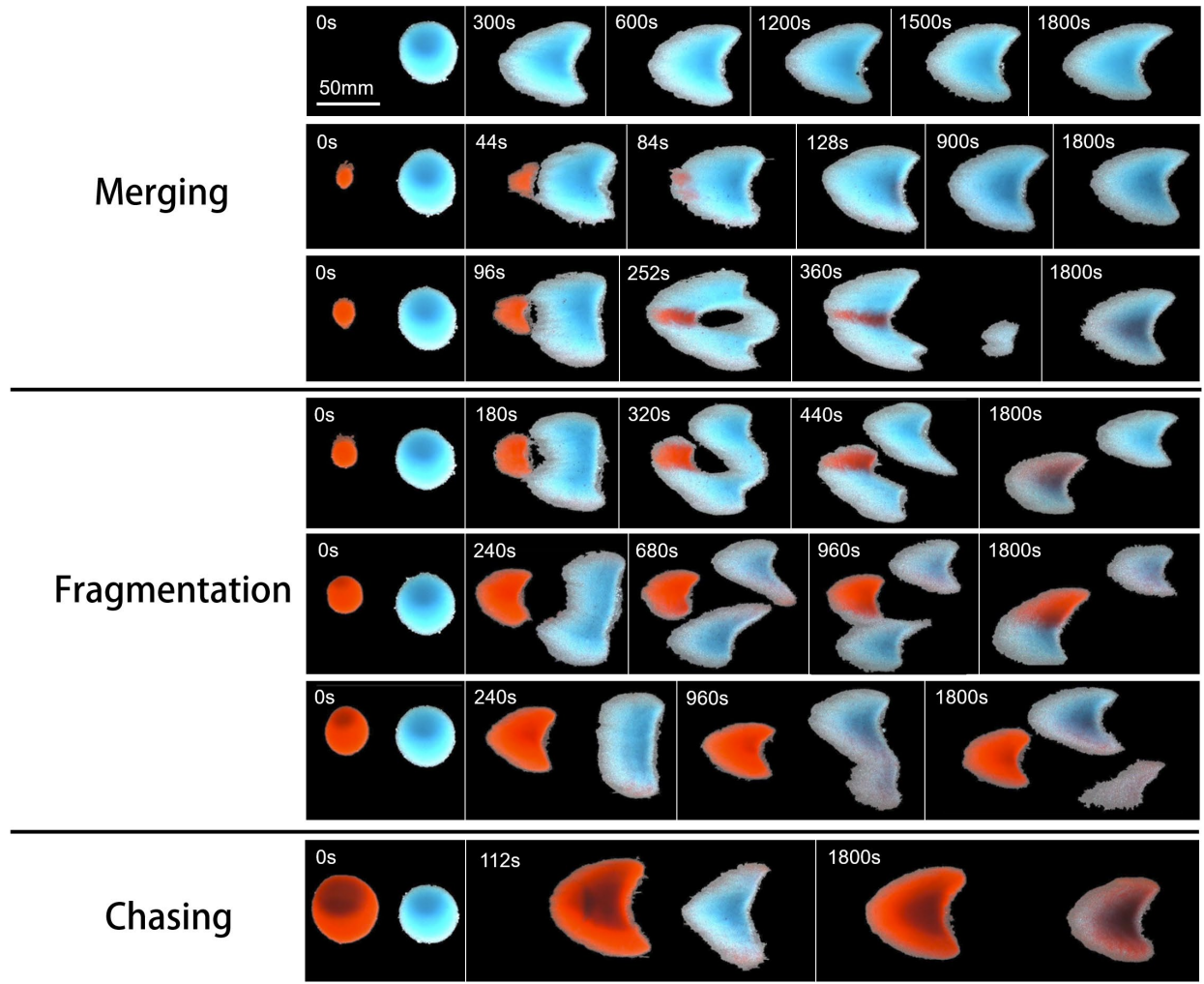

**Figure S4.** Snapshots from the bottom view of dune collision with the increase in the initial mass ratio ($V_f$ = 0.21 m/s). The patterns are as follows: isolated case ($\xi_m = 0.00$), merging ($\xi_m = 0.02$), exchange ($\xi_m = 0.05$), fragmentation-exchange ($\xi_m = 0.09$), exchange-chasing ($\xi_m = 0.20$), fragmentation-chasing ($\xi_m = 0.35$), and chasing ($\xi_m = 1.50$).

**Table S1.** List of tests. $m_R$ and $m_B$ are the mass of the upstream and downstream dunes, respectively, $\xi_m = m_R / m_B$ is the initial mass ratio, $V_f$ is the mean rate of water flow, $\varepsilon_x$ and $\varepsilon_y$ are the initial distance of the centroid of two dunes in the flowing and transverse directions, respectively, $R_R$ and $R_B$ are the initial radius of the upstream and downstream dunes, respectively.

| # | $m_R$ | $m_B$ | $\xi_m$ | $V_f$ | $\varepsilon_x$ | $\varepsilon_y$ | $\varepsilon_x/R_R$ | $R_R$ | $R_B$ | *Pattern* |
|---|---|---|---|---|---|---|---|---|---|---|
| # | (g) | (g) | — | m/s | cm | cm | — | cm | cm | — |
| 1 | 0.00 | 10.00 | 0.00 | 0.19 | — | — | — | — | 2.32 | Isolated case |
| 2 | 0.10 | 10.00 | 0.01 | 0.19 | 5.31 | 0.22 | 6.82 | 0.78 | 2.26 | Merging |
| 3 | 0.20 | 10.00 | 0.02 | 0.19 | 5.40 | 0.51 | 6.59 | 0.82 | 2.36 | Merging |
| 4 | 0.30 | 10.00 | 0.03 | 0.19 | 5.67 | 0.64 | 6.19 | 0.92 | 2.39 | Merging |
| 5 | 0.40 | 10.00 | 0.04 | 0.19 | 6.11 | 0.54 | 5.91 | 1.03 | 2.34 | Exchange |
| 6 | 0.50 | 10.00 | 0.05 | 0.19 | 6.29 | 0.54 | 6.30 | 1.00 | 2.27 | Exchange |
| 7 | 0.70 | 10.00 | 0.07 | 0.19 | 6.27 | 0.13 | 6.01 | 1.04 | 2.31 | Exchange |
| 8 | 0.90 | 10.00 | 0.09 | 0.19 | 5.76 | 0.41 | 4.43 | 1.30 | 2.31 | Frag-exchange |
| 9 | 1.00 | 10.00 | 0.10 | 0.19 | 6.61 | 0.20 | 5.84 | 1.13 | 2.33 | Frag-exchange |
| 10 | 1.20 | 10.00 | 0.12 | 0.19 | 7.07 | 0.55 | 5.60 | 1.26 | 2.30 | Frag-exchange |
| 11 | 1.40 | 10.00 | 0.14 | 0.19 | 6.56 | 0.47 | 4.88 | 1.34 | 2.28 | Frag-exchange |
| 12 | 1.70 | 10.00 | 0.17 | 0.19 | 6.32 | 0.60 | 4.41 | 1.43 | 2.28 | Frag-exchange |
| 13 | 2.00 | 10.00 | 0.20 | 0.19 | 6.16 | 0.76 | 4.00 | 1.54 | 2.28 | Chasing-exchange |
| 14 | 2.30 | 10.00 | 0.23 | 0.19 | 6.25 | 0.39 | 4.15 | 1.51 | 2.24 | Chasing-exchange |
| 15 | 2.50 | 10.00 | 0.25 | 0.19 | 6.15 | 0.50 | 3.94 | 1.56 | 2.27 | Chasing-exchange |
| 16 | 2.70 | 10.00 | 0.27 | 0.19 | 6.32 | 0.21 | 3.98 | 1.59 | 2.32 | Frag-chasing |
| 17 | 3.00 | 10.00 | 0.30 | 0.19 | 6.77 | 0.15 | 3.88 | 1.75 | 2.35 | Frag-chasing |
| 18 | 3.20 | 10.00 | 0.32 | 0.19 | 6.24 | 0.34 | 3.76 | 1.66 | 2.35 | Chasing-exchange |
| 19 | 3.50 | 10.00 | 0.35 | 0.19 | 6.06 | 0.01 | 3.50 | 1.73 | 2.42 | Frag-chasing |
| 20 | 3.80 | 10.00 | 0.38 | 0.19 | 6.68 | 0.37 | 3.77 | 1.77 | 2.33 | Frag-chasing |
| 21 | 4.00 | 10.00 | 0.40 | 0.19 | 6.35 | 0.26 | 3.54 | 1.80 | 2.32 | Frag-chasing |
| 22 | 4.20 | 10.00 | 0.42 | 0.19 | 6.14 | 0.13 | 3.45 | 1.78 | 2.26 | Frag-chasing |
| 23 | 4.50 | 10.00 | 0.45 | 0.19 | 6.01 | 0.03 | 3.25 | 1.85 | 2.28 | Frag-chasing |
| 24 | 4.70 | 10.00 | 0.47 | 0.19 | 6.26 | 0.07 | 3.46 | 1.81 | 2.29 | Frag-chasing |
| 25 | 5.30 | 10.00 | 0.53 | 0.19 | 6.71 | 0.19 | 3.44 | 1.95 | 2.27 | Frag-chasing |
| 26 | 5.50 | 10.00 | 0.55 | 0.19 | 6.78 | 0.46 | 3.40 | 1.99 | 2.29 | Chasing |
| 27 | 5.70 | 10.00 | 0.57 | 0.19 | 6.78 | 0.27 | 3.47 | 1.96 | 2.28 | Frag-chasing |
| 28 | 6.00 | 10.00 | 0.60 | 0.19 | 6.59 | 0.28 | 3.29 | 2.00 | 2.27 | Chasing |
| 29 | 6.30 | 10.00 | 0.63 | 0.19 | 7.36 | 0.35 | 3.57 | 2.06 | 2.27 | Chasing |
| 30 | 7.50 | 10.00 | 0.75 | 0.19 | 6.62 | 0.33 | 3.09 | 2.15 | 2.33 | Chasing |
| 31 | 8.00 | 10.00 | 0.80 | 0.19 | 7.66 | 0.43 | 3.47 | 2.21 | 2.33 | Chasing |
| 32 | 10.00 | 10.00 | 1.00 | 0.19 | 7.04 | 0.03 | 2.99 | 2.35 | 2.24 | Chasing |
| 33 | 12.00 | 10.00 | 1.20 | 0.19 | 7.22 | 0.41 | 2.89 | 2.50 | 2.29 | Chasing |
| 34 | 13.00 | 10.00 | 1.30 | 0.19 | 6.71 | 0.26 | 2.58 | 2.60 | 2.37 | Chasing |
| 35 | 15.00 | 10.00 | 1.50 | 0.19 | 6.95 | 0.50 | 2.55 | 2.72 | 2.40 | Chasing |
| 36 | 0.00 | 10.00 | 0.00 | 0.21 | — | — | — | — | 2.35 | isolated case |
| 37 | 0.20 | 10.00 | 0.02 | 0.21 | 6.44 | 0.38 | 8.87 | 0.73 | 2.38 | Merging |
| 38 | 0.50 | 10.00 | 0.05 | 0.21 | 6.47 | 0.43 | 6.74 | 0.96 | 2.34 | Exchange |

| 39 | 0.70 | 10.00 | 0.07 | 0.21 | 6.25 | 0.44 | 5.88 | 1.06 | 2.33 | Frag-exchange |
| 40 | 0.90 | 10.00 | 0.09 | 0.21 | 6.65 | 0.76 | 5.65 | 1.18 | 2.35 | Frag-exchange |
| 41 | 1.20 | 10.00 | 0.12 | 0.21 | 6.46 | 0.46 | 5.34 | 1.21 | 2.31 | Frag-exchange |
| 42 | 1.50 | 10.00 | 0.15 | 0.21 | 6.40 | 0.42 | 4.57 | 1.40 | 2.35 | Chasing-exchange |
| 43 | 1.70 | 10.00 | 0.17 | 0.21 | 5.89 | 0.49 | 4.39 | 1.34 | 2.38 | Frag-exchange |
| 44 | 2.00 | 10.00 | 0.20 | 0.21 | 6.26 | 0.69 | 4.47 | 1.40 | 2.32 | Chasing-exchange |
| 45 | 2.30 | 10.00 | 0.23 | 0.21 | 6.10 | 0.40 | 4.12 | 1.48 | 2.34 | Chasing-exchange |
| 46 | 2.50 | 10.00 | 0.25 | 0.21 | 6.33 | 0.44 | 4.04 | 1.57 | 2.29 | Chasing-exchange |
| 47 | 2.70 | 10.00 | 0.27 | 0.21 | 6.53 | 0.43 | 4.17 | 1.57 | 2.29 | Frag-chasing |
| 48 | 3.00 | 10.00 | 0.30 | 0.21 | 6.20 | 0.33 | 3.85 | 1.61 | 2.30 | Chasing-exchange |
| 49 | 3.50 | 10.00 | 0.35 | 0.21 | 6.14 | 0.38 | 3.51 | 1.75 | 2.34 | Frag-chasing |
| 50 | 4.00 | 10.00 | 0.40 | 0.21 | 6.15 | 0.06 | 3.46 | 1.78 | 2.30 | Chasing |
| 51 | 4.50 | 10.00 | 0.45 | 0.21 | 6.33 | 0.42 | 3.41 | 1.85 | 2.33 | Chasing |
| 52 | 5.00 | 10.00 | 0.50 | 0.21 | 5.95 | 0.32 | 3.13 | 1.90 | 2.30 | Frag-chasing |
| 53 | 5.50 | 10.00 | 0.55 | 0.21 | 6.08 | 0.24 | 2.97 | 2.04 | 2.33 | Chasing |
| 54 | 6.00 | 10.00 | 0.60 | 0.21 | 6.31 | 0.37 | 3.16 | 2.00 | 2.28 | Chasing |
| 55 | 6.50 | 10.00 | 0.65 | 0.21 | 6.19 | 0.28 | 3.07 | 2.02 | 2.40 | Chasing |
| 56 | 7.00 | 10.00 | 0.70 | 0.21 | 6.70 | 0.24 | 3.22 | 2.08 | 2.31 | Chasing |
| 57 | 7.50 | 10.00 | 0.75 | 0.21 | 6.22 | 0.02 | 2.99 | 2.08 | 2.29 | Chasing |
| 58 | 8.00 | 10.00 | 0.80 | 0.21 | 6.19 | 0.03 | 2.95 | 2.10 | 2.29 | Chasing |
| 59 | 9.00 | 10.00 | 0.90 | 0.21 | 6.18 | 0.05 | 2.77 | 2.23 | 2.29 | Chasing |
| 60 | 10.00 | 10.00 | 1.00 | 0.21 | 6.34 | 0.12 | 2.72 | 2.33 | 2.32 | Chasing |
| 61 | 12.00 | 10.00 | 1.20 | 0.21 | 6.01 | 0.04 | 2.47 | 2.43 | 2.30 | Chasing |
| 62 | 13.00 | 10.00 | 1.30 | 0.21 | 6.23 | 0.07 | 2.44 | 2.55 | 2.30 | Chasing |
| 63 | 15.00 | 10.00 | 1.50 | 0.21 | 6.46 | 0.38 | 2.45 | 2.63 | 2.29 | Chasing |
| 64 | 0.00 | 10.00 | 0.00 | 0.23 | — | — | — | — | 2.32 | isolated case |
| 65 | 0.20 | 10.00 | 0.02 | 0.23 | 6.72 | 0.29 | 10.13 | 0.66 | 2.28 | Merging |
| 66 | 0.50 | 10.00 | 0.05 | 0.23 | 6.83 | 0.10 | 8.11 | 0.84 | 2.31 | Merging |
| 67 | 0.60 | 10.00 | 0.06 | 0.23 | 6.44 | 0.51 | 7.29 | 0.88 | 2.32 | Exchange |
| 68 | 0.70 | 10.00 | 0.07 | 0.23 | 6.80 | 0.40 | 6.70 | 1.01 | 2.30 | Exchange |
| 69 | 0.90 | 10.00 | 0.09 | 0.23 | 6.72 | 0.06 | 6.05 | 1.11 | 2.32 | Exchange |
| 70 | 1.20 | 10.00 | 0.12 | 0.23 | 6.46 | 0.40 | 5.35 | 1.21 | 2.26 | Exchange |
| 71 | 1.30 | 10.00 | 0.13 | 0.23 | 6.38 | 0.48 | 5.23 | 1.22 | 2.33 | Frag-exchange |
| 72 | 1.50 | 10.00 | 0.15 | 0.23 | 6.88 | 0.30 | 5.36 | 1.28 | 2.33 | Exchange |
| 73 | 1.70 | 10.00 | 0.17 | 0.23 | 6.44 | 0.08 | 4.71 | 1.37 | 2.31 | Exchange |
| 74 | 2.00 | 10.00 | 0.20 | 0.23 | 6.70 | 0.40 | 4.58 | 1.46 | 2.32 | Frag-exchange |
| 75 | 2.30 | 10.00 | 0.23 | 0.23 | 6.59 | 0.13 | 4.25 | 1.55 | 2.32 | Frag-exchange |
| 76 | 2.50 | 10.00 | 0.25 | 0.23 | 6.44 | 0.26 | 4.39 | 1.47 | 2.29 | Chasing-exchange |
| 77 | 2.70 | 10.00 | 0.27 | 0.23 | 6.06 | 0.01 | 3.70 | 1.64 | 2.41 | Frag-chasing |
| 78 | 3.00 | 10.00 | 0.30 | 0.23 | 5.97 | 0.08 | 3.73 | 1.60 | 2.52 | Chasing-exchange |
| 79 | 3.50 | 10.00 | 0.35 | 0.23 | 6.68 | 0.20 | 4.00 | 1.67 | 2.34 | Frag-chasing |
| 80 | 4.00 | 10.00 | 0.40 | 0.23 | 6.20 | 0.10 | 3.59 | 1.73 | 2.34 | Frag-chasing |
| 81 | 4.50 | 10.00 | 0.45 | 0.23 | 6.48 | 0.16 | 3.57 | 1.81 | 2.37 | Chasing |
| 82 | 5.00 | 10.00 | 0.50 | 0.23 | 6.55 | 0.15 | 3.60 | 1.82 | 2.29 | Frag-chasing |
| 83 | 6.00 | 10.00 | 0.60 | 0.23 | 6.24 | 0.30 | 3.20 | 1.95 | 2.25 | Chasing |
| 84 | 6.50 | 10.00 | 0.65 | 0.23 | 6.26 | 0.29 | 3.15 | 1.99 | 2.31 | Chasing |
| 85 | 7.50 | 10.00 | 0.75 | 0.23 | 6.26 | 0.17 | 2.93 | 2.14 | 2.27 | Chasing |
| 86 | 8.00 | 10.00 | 0.80 | 0.23 | 6.19 | 0.45 | 2.87 | 2.16 | 2.28 | Chasing |
| 87 | 10.00 | 10.00 | 1.00 | 0.23 | 6.46 | 0.19 | 2.81 | 2.30 | 2.24 | Chasing |
| 88 | 12.00 | 10.00 | 1.20 | 0.23 | 6.61 | 0.05 | 2.70 | 2.45 | 2.28 | Chasing |
| 89 | 13.00 | 10.00 | 1.30 | 0.23 | 6.26 | 0.14 | 2.43 | 2.57 | 2.34 | Chasing |

| 90 | 15.00 | 10.00 | 1.50 | 0.23 | 6.20 | 0.27 | 2.30 | 2.69 | 2.25 | Chasing |
| 91 | 17.00 | 10.00 | 1.70 | 0.24 | 6.07 | 0.32 | 2.22 | 2.74 | 2.27 | "Soliton" |
| 92 | 17.00 | 10.00 | 1.70 | 0.24 | 6.63 | 0.50 | 2.44 | 2.71 | 2.28 | "Soliton" |
| 93 | 17.00 | 10.00 | 1.70 | 0.24 | 7.39 | 0.37 | 2.70 | 2.74 | 2.33 | "Like-soliton" |
| 94 | 17.00 | 10.00 | 1.70 | 0.24 | 8.17 | 0.12 | 2.99 | 2.73 | 2.24 | "Like-soliton" |
| 95 | 17.00 | 10.00 | 1.70 | 0.24 | 9.00 | 0.07 | 3.26 | 2.76 | 2.30 | Chasing |

**LBM simulation**

The LBM code for simulating the flow around the two dunes shown in Fig. 6 is modified according to the source code for 2D cavity flow from Guo and Shu (2013). The process of verifying the validity of the code by simulating the flow around a cylinder (Guo & Shu, 2013). Fig. S5 shows the flow domain and boundary conditions. The ratios of the wake size to the diameter of the cylinder and the drag coefficients $C_d$ of the flow direction obtained by the simulation are very close to the results of the existing data (Nieuwstadt & Keller, 1973; Coutanceau & Bouard, 1977; He & Doolen, 1997). Therefore, it can be concluded that the code is verified.

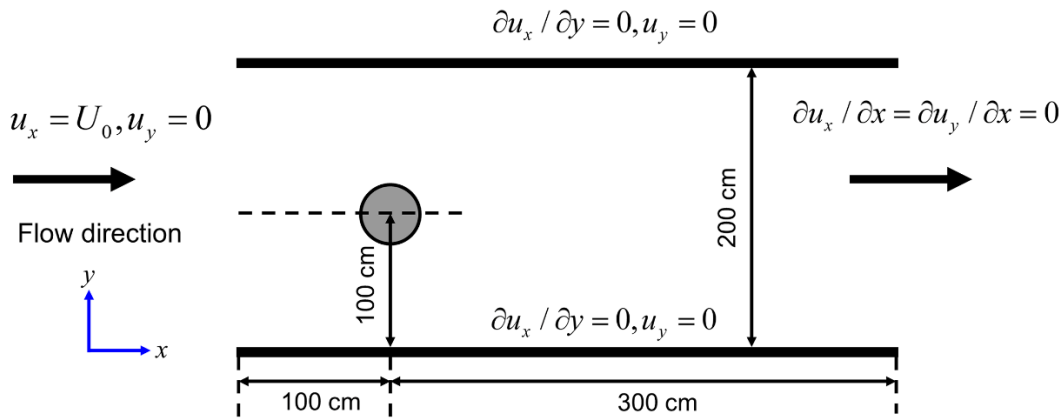

**Figure. S5** Sketch of the flow around a cylinder

The flow domain for simulating the dune interaction is $N_x \times N_y$=2000 cm×1000 cm. The velocity in the inlet $U_0$=20 cm/s corresponds to the motion (speed) boundary scheme. The flow in the outlet is fully developed. The side and cylindrical boundaries execute the standard bounce-back scheme. Via planar PIV measurement, Bristow et al. (2018, 2019) studied in a decoupled manner the turbulent flow around fixed barchan models in a series of configurations mimicking the early stage of a laterally offset collision between the two dunes, which makes it reasonable that the transient dune contours at $t$=175 s under $\xi_m$=0.38, $\xi_m$=1.50 and $\xi_m$=1.70 in the present experiment were used as obstacles to study the flow structure around dune body. The original images were scaled by a factor of 0.2, with the center of the upstream barchan placed at (400 cm, 500 cm).

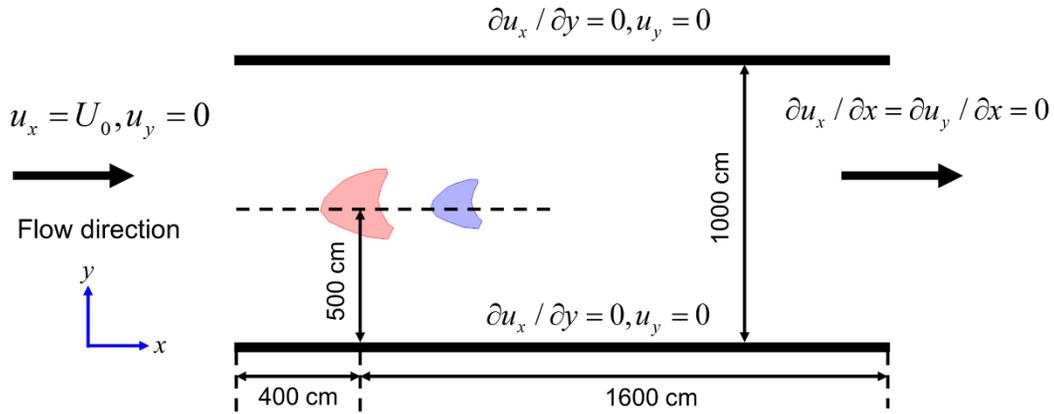

**Figure. S6** Sketch of the flow around dual barchans.

The $\Omega$-criterion (Hunt et al., 1988; Liu et al., 2016) was used to extract the vortex structure in the flow field, where the threshold $\Omega$=0.52 was applied, and the expression of $\Omega$ was given as

$$\Omega = \frac{\| \boldsymbol{B} \|_F^2}{\| \boldsymbol{A} \|_F^2 + \| \boldsymbol{B} \|_F^2 + \varepsilon} \tag{1}$$

where $\| \ \|_F$ is the Frobenius norm of the matrix. The symmetric tensor $\boldsymbol{A}$ has the effect of canceling the antisymmetric tensor $\boldsymbol{B}$ acting on the rigid body to make it rotate. Therefore, $\Omega$ is introduced to represent the ratio of vortical vorticity over the whole vorticity inside a vortex core. A small positive number $\varepsilon$ is used to make sure the denominator is not zero. Dong et al. (2018) proposed a method for selecting $\varepsilon$, which avoids artificial selection of $\varepsilon$, and has achieved good results in practical applications:

$$\varepsilon = 0.001 \times (\| \boldsymbol{B} \|_F^2 - \| \boldsymbol{A} \|_F^2).$$ (2)

Taking $\xi_m = 1.70$ as an example, Fig. S7 shows the LBM simulation results of the vortex shedding and the evolution of the interdune vortex between 150 and 200 seconds.

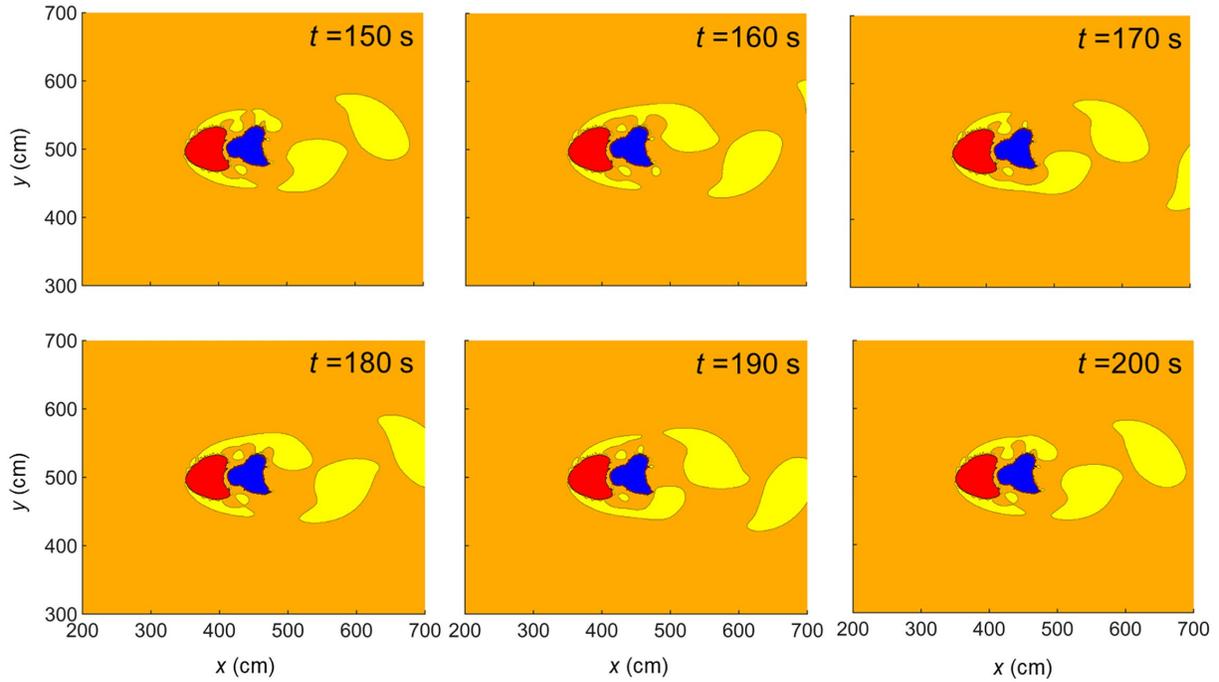

**Figure. S7** LBM simulation results ($\xi_m$=1.70)